\documentclass[12pt]{article}

\newcommand{\be}{\begin{equation}}
 \newcommand{\ee}{\end{equation}}

\begin{document}
\vspace*{-.6in} \thispagestyle{empty}
\begin{flushright}
gr-qc/0011089
\end{flushright}
\baselineskip = 20pt

\vspace{.5in} {\Large
\begin{center}
Quantum Gravity at the Turn of the Millennium\end{center}}

\begin{center}
Gary T. Horowitz
\\ \emph{Physics Department\\
University of California\\
Santa Barbara, CA  93111 USA}
\end{center}
\vspace{1in}

\begin{center}
\textbf{Abstract}
\end{center}
\begin{quotation}
\noindent 
A very  brief review  is given of the current state of research in quantum 
gravity. Over
the past fifteen years, two approaches have emerged as the most
promising  paths to  a quantum theory of gravity:
string theory and quantum geometry. I will discuss
the main achievements and open problems of each of these approaches,
and compare their strengths and weaknesses.
\end{quotation}
\vspace{1in}

\centerline{Talk presented at the {\it Ninth Marcel Grossmann Meeting} (MG9)
Rome, July, 2000}
\vfil

\newpage

\pagenumbering{arabic} 

\section{Introduction}

Two of the greatest achievements of physics in the last century were Einstein's
general theory of relativity and quantum theory. Each of these theories
has been extremely well tested and has been very successful. However, they
are mutually incompatible. Thus our basic understanding of nature is not only
incomplete -- but inconsistent. One clearly needs a new theory, quantum gravity,
which incorporates the principles of both of these theories and reduces to them
in appropriate limits.

At first sight, the problem of constructing a quantum theory of gravity sounds
easy since there are no experimental constraints! The task is simply to find any
theory which unifies  general relativity and quantum theory. However, on 
second thought, the problem sounds extremely difficult. General relativity
teaches us that gravity is just a manifestation of the curvature of space and
time. So quantum gravity must involve the quantization of space and time,
something we have no previous experience with.

Surprisingly, even though there are no experimental constraints, this is a 
constraint on quantum gravity which was found in the early 1970's by studying
black holes. Motivated by the close analogy between the laws of  black hole
mechanics and ordinary thermodynamics, Bekenstein  proposed 
that black holes
have an entropy proportional to their horizon area $A$ \cite{BEK}.
Then Hawking showed that
if matter is treated quantum mechanically (but gravity remains classical),
black holes emit thermal radiation with a temperature $T = \hbar \kappa/2\pi$
where $\kappa$ is the surface gravity of the black hole \cite{HAW}. 
This confirmed 
Bekenstein's ideas and fixed the coefficient:
\be
S_{bh} = {A\over 4 G \hbar}
\ee
This is an enormous entropy -- much larger than the entropy in the matter that
collapsed to form the black hole.
In a more fundamental statistical description, the 
entropy should be a measure of the log of the number of accessible states.
So a constraint on any candidate quantum theory of gravity is to show
that the number of quantum states associated with a black hole is indeed
$e^{S_{bh}}$. We will see that there has been considerable progress in 
satisfying this constraint recently.

Over the years, there has been much discussion of possible consequences of 
quantum gravity. Let me comment on a few of the most popular:

$\bullet$ {\it Quantum gravity will smooth out spacetime singularities}

This is false as stated. Quantum gravity cannot smooth out all singularities.
Some timelike singularities must remain, such as the one in the $M<0$
Schwarzschild solution. This can be seen from the following simple argument
\cite{HOMY}. Since quantum gravity must reduce to general relativity for 
weak fields (and large number of quanta), there must be solutions which look
like $M<0$ Schwarzschild at large radii, for any $M$.
At small radii, the
curvature becomes strong and the solution may be significantly modified. But
if it is not singular in some sense, it would represent a state  in the
theory with negative total energy. Since the energy is unbounded from below,
there would be no stable vacuum state.

$\bullet$ {\it Quantum gravity will allow the topology of space to change}

This is almost certainly true. There are semi-classical calculations of
pair creation of magnetically charged black holes in a background
magnetic field \cite{GAST}. One can reliably calculate this rate for black holes
with size much larger than the Planck scale. It is extremely small,
but it does change the topology of space from $R^3$ to $R^3$ with an 
$S^2 \times S^1$ wormhole attached. This is because the black holes are
created with their horizons identified. It is interesting to note that if one
compares the rate of black hole pair creation to the rate of magnetic
monopole creation, one sees an enhancement in the black hole case of
$e^{S_{bh}}$ in line with the expectation that there are $e^{S_{bh}}$
different species of black holes \cite{GGS}.
We will see further evidence for topology 
change shortly.

$\bullet$ {\it Black hole evaporation will violate quantum mechanics: pure
states will evolve to mixed states}

Since black holes can be formed from matter in a pure state and the radiation
emitted is thermal (in the  semiclassical approximation) it was
thought that black hole evaporation would cause pure states to evolve into
mixed states.
There is recent evidence (discussed below) that this is false. In a 
more exact treatment there are likely
to be correlations between the radiation emitted at early time and later
time  which ensure that the evolution is unitary.

$\bullet$ {\it  Space and time will not be fundamental, but derived properties.}

This is likely to be true,  but the key question is, what replaces them?

Over the past fifteen years, there has been significant progress in two
different approaches to quantum gravity: string theory and quantum geometry.
String theory attempts to provide a unified theory of all forces and particles
as well as a quantum theory of gravity.
Quantum geometry attempts to quantize general relativity (by itself)
in a background independent, non-perturbative way. 
It is well known that the usual perturbative approach to
quantizing general relativity fails since the theory
is not renormalizable. But quantum general relativity may still
exist nonperturbatively.
Due to lack of time, I will not attempt to describe these approaches
in detail. (The details keep changing anyway.) Instead, I will try to 
summarize the current status of each approach and some of the results that
have been achieved so far.

Note on references: The list of references is necessarily incomplete and
somewhat subjective. I have tried to include at least some of the
key papers describing recent developments in each approach.

\section{String theory}

String theory\footnote{A good general reference for string theory including
many of the results
discussed in this section is \cite{POL}.} starts with the idea that
all elementary particles are not pointlike, but excitations of a one dimensional
string. If one quantizes a 
free relativistic string in flat
spacetime one finds a infinite tower of modes of increasing mass. There is
a massless spin two mode which is identified with a linearized graviton.
Next, one postulates a simple splitting and joining interaction between strings
and finds, remarkably,
that this reproduces the perturbative expansion of general
relativity. One then adds fermionic degrees of freedom to the string so the
theory is supersymmetric. This makes the theory better behaved and calculations
easier to control. 
The consistent quantization of a string turns out to impose a 
constraint on the spacetime dimension. In most cases, one needs ten spacetime
dimensions. Contact with observations obviously
requires that six of these dimensions are unobservable. The simplest possibility
is the old Kaluza-Klein idea that these six dimensions are wrapped up in a 
small compact manifold\footnote{Other possibilities are also being explored
\cite{ADD,RS}.}. The natural size of this compact manifold is the string length,
a new dimensionful parameter in the theory set by the string tension. The
string length, $\ell_s$, is related to the Planck length, $\ell_p$, by a power
of the (dimensionless) string coupling $g$. In ten dimensions, $\ell_p =
g^{1/4} \ell_s$.

Since the string represents fluctuations about the background spacetime
it is propagating in, the above description is strictly perturbative.
String theory has now progressed far beyond  this 
perturbative beginning.
The current status of string theory is roughly the following.

\begin{enumerate}
\item Classical theory is well understood. The classical equations resemble
general relativity with an infinite series of correction terms involving
higher powers and derivatives of the curvature. Since the correction
terms become important only when the curvature is of order the string scale,
any solution to general relativity with curvature smaller than $\ell^{-2}_s$
is an approximate classical solution to  string theory.
To obtain exact solutions, it is useful to note that the classical
field equations arise from the vanishing of the conformal anomaly in a 
certain two dimensional field theory. Several classes
of exact solutions are known (often using special properties to ensure that the
higher order correction terms vanish). One class, based on supersymmetry, 
includes
compact Ricci flat six manifolds known as Calabi-Yau spaces which can be 
used to compactify spacetime down to four dimensions. Other classes include
exact plane waves and group manifolds.

\item Quantum perturbation  theory is well understood and well behaved.
In particular, it is finite order by order in the loop expansion. Thus
string theory provides a perturbatively finite quantum theory of gravity.
However, the perturbation theory does not determine the quantum theory
uniquely. Nonperturbative effects are important. This is evident in the
fact that the loop expansion does not converge.

\item Some nonperturbative properties are known. 
These are mostly through clever 
use of supersymmetry, which guarantees that certain properties which
are valid at weak coupling must continue to hold at strong coupling. 
In particular, extended
objects (called branes) play an important role in the theory. At the
perturbative level, there are five different string theories in ten dimensions
which differ in the amount of supersymmetry and fundamental gauge groups they
contain. There is convincing evidence that the strong coupling limit of one
theory is equivalent to the weak coupling limit of another theory. These are
known as duality symmetries. In fact, it is now believed that all of these
theories can be obtained from a single eleven dimensional theory\footnote{This
is often called M-theory, but I will continue to refer to this approach as
string theory.}.

\item There exists a complete nonperturbative formulation of the theory 
for certain boundary conditions. This will be discussed further below.

\end{enumerate}

Up until five years ago, the status of string theory was essentially just
the first two points above. Point three has an interesting consequence.
Since string theory includes gravity and strong
coupling implies strong gravitational fields, one might have expected
the strong coupling limit of string theory 
to have large fluctuations of space and time, i.e.,  spacetime foam.
But this does not seem to be the case.
It appears that the strong coupling limit of
the theory can be described in terms of a weakly coupled theory in 
new variables. There is no evidence for spacetime foam.

Now I turn to discuss some results in string theory. 

A) {\it Singularities:} The first thing to note is that the definition of a 
singularity is different in string theory than in general relativity, even
classically. In general relativity, we usually  define a singularity in
terms of geodesic incompleteness which is based on the motion of test particles.
In string theory, we must use test strings. So a spacetime is considered 
singular if test strings are not well behaved\footnote{Strictly speaking,
one should also require that the other extended objects 
in the theory --branes -- have well behaved propagation.}. It turns out
that some spacetimes which are singular in general relativity are
completely nonsingular in string theory. A simple example is the quotient
of Euclidean space by a discrete subgroup of the rotation group. The
resulting space, called an orbifold, has a conical singularity at the
origin. Even though this leads to geodesic incompleteness in general 
relativity, it is completely harmless in string theory. This is essentially
because strings are extended objects. 

The orbifold has a very mild singularity,
but even curvature singularities can be harmless in string theory. 
As mentioned above, string theory has exact solutions which are the product
of four dimensional Minkowski space, and a compact Calabi-Yau
space. A given Calabi-Yau manifold usually admits a whole family of 
Ricci flat metrics. So one can construct a solution in which the four large
dimensions stay approximately flat and the geometry of the Calabi-Yau 
manifold changes slowly from one Ricci flat metric to another. In this process
the  Calabi-Yau space can develop a curvature singularity.
In many cases, this can be viewed as arising from 
a topologically nontrivial $S^2$ or $S^3$ being shrunk down
to zero area. 
It has been shown that when this happens, string theory remains completely well
defined. The evolution continues through the geometrical
singularity to a nonsingular Calabi-Yau space on the other side 
\cite{AGM,STR}.

The reason this happens is roughly the following. There are extra degrees of
freedom in the theory
associated with branes wrapped around topologically 
nontrivial surfaces. As long as the area of the surface is nonzero, these
degrees of freedom are massive, and it is consistent to ignore them. However
when the surface shrinks to zero volume these degrees of freedom become
massless, and one must include them in the analysis. When this is done, 
the theory is nonsingular.

The above singularities are all in the extra spatial dimensions. However
other singularities which involve time in a crucial way have also been shown
to be harmless. Putting many branes on top of each other produces a 
gravitational field which often has a curvature singularity at the location
of the brane.  It has been shown that one can understand
physical processes near this
singularity 
in terms of excitations of the branes.

Despite all this progress,
we still do not yet have an understanding of the most important
types of  singularities: those arising from gravitational collapse
or cosmology.

B) {\it Topology change:} It has been shown unambiguously that the topology
of space can change in string theory.
In fact, when one evolves through
a singular Calabi-Yau space as described above, the topology of the manifold
changes \cite{AGM,AMS}.
A simpler example of topology change is the following. Consider
one direction in  space compactified to a circle. If one identifies points
under a shift $\theta \rightarrow \theta +\pi$, one obtains a circle of
half the radius. If one identifies points under a reflection about a
diameter, one obtains a line segment. It turns out that for a circle
whose radius is the string scale, one can show 
these two $Z_2$ actions are equivalent in string theory\footnote{Usually,
compactifying on a circle produces a $U(1)$ gauge field. At this special
radius, there is
an enhanced $SU(2)$ symmetry and these two $Z_2$ actions are conjugate
subgroups.}.  There is no way for
strings to distinguish them. So one can start with one direction compactified on
a large circle, slowly shrink it down to the string scale, replace it with
a line segment, and then slowly expand the line segment. As far as
strings are concerned, the evolution is completely nonsingular.

C) {\it Black hole entropy:} By far the most important result is that
it has been shown that string theory can satisfy the black hole constraint
mentioned earlier. For a large class of extreme and near extreme 
charged black holes,
one can count the number of quantum string states at weak coupling
with the same mass and
charge as the black hole. The answer turns out to  agree
exactly with the prediction made by Bekenstein and Hawking \cite{STVA,REVS}.
It is important
to note that it is not just one number being reproduced. One
can consider black holes with several different types of charges and
angular momentum. String theory correctly reproduces the entropy as a function
of all of these parameters. 

The calculations are quite remarkable since
one starts at weak coupling where gravitational effects are turned off and
spacetime is flat. One considers configurations of branes and strings with
appropriate charges and counts the number of states with a given energy.
One then increases the string coupling. The gravitational field of the
branes and strings becomes stronger and they eventually form a black hole.
One compares the Bekenstein-Hawking entropy of the resulting black hole
and finds complete agreement with the log of the number of states computed
in flat spacetime. It is possible to do this calculation exactly only for
extremal and near extremal black holes. For more general black holes, one
can show that the log of the number of string states is proportional to the
area \cite{HOPO}, but the coefficient is difficult to calculate.

In some cases, one can calculate the spectrum of Hawking radiation  in string
theory and show that it agrees with the semiclassical calculation.
This is
remarkable since the spectrum is not exactly thermal, but has grey body
factors arising from spacetime curvature outside the horizon. These are
correctly reproduced in string  theory, even though the string calculation is
done in flat space. The calculations look completely different, but the
results agree.

D) {\it Nonperturbative formulation:} By studying these black hole results
more closely, people were led to a new and more complete formulation of the
theory. 
\vskip .5cm

{\bf AdS/CFT Conjecture} (Maldacena \cite{MAL}): 
String theory on spacetimes which 
asymptotically approach the product of anti de Sitter (AdS) and a compact
space, is completely described by a conformal field theory (CFT)
``living on the
boundary at infinity".

\vskip .5cm

In particular, string theory with $AdS_5\times S^5$ boundary conditions
is described by a four dimensional supersymmetric $SU(N)$ gauge theory.
Since the gauge theory is defined nonperturbatively,
this is a nonperturbative and (mostly) background independent
formulation of string theory. A background spacetime metric only enters
in the boundary conditions at infinity.

At first sight this conjecture seems unbelievable. 
How could an ordinary field theory
describe all of string theory? I don't have time to describe the impressive
body of evidence in favor of this correspondence which has accumulated over
the past few years. In the past three years, more than a thousand
papers have been written on various aspects of this conjecture. A good
review is \cite{MAGOO}.

This conjecture provides a ``holographic" description of quantum gravity
in that the fundamental degrees of freedom live on a lower dimensional
space. The idea that quantum gravity might be holographic 
was first suggested by 't Hooft \cite{HOO} and Susskind \cite{SUS} motivated by
the fact that black hole entropy is proportional to its horizon area. 
It also confirms earlier indications that string theory has fewer 
fundamental degrees of
freedom than it appears in perturbation theory. This conjecture
provides an answer to the longstanding question raised in the introduction:
If space and time are not fundamental, what replaces them? Here the answer
is that there is an auxiliary spacetime metric which is fixed by the 
boundary conditions at infinity. The CFT uses this metric, but the physical
spacetime metric is a derived quantity. The dictionary relating spacetime
concepts in the bulk and field theory concepts on the boundary
is very incomplete, and still being 
developed.

This conjecture has an interesting consequence. Consider the formation
and evaporation of a small black hole in a spacetime which is asymptotically
$AdS_5\times S^5$. By the AdS/CFT correspondence, this process is described
by ordinary unitary evolution in the CFT. So black hole evaporation 
does not violate quantum mechanics. This is the basis for my  earlier 
comment that the belief that black hole evaporation is not unitary
is probably false.

\section{Quantum Geometry}

Quantum geometry\footnote{For a general review of this approach, 
see \cite{ROV}.}
begins by rewriting (four dimensional)
general relativity in first order form in terms of a tetrad and connection.
But instead of the
usual Lorentz connection, one uses a self dual connection.
One then casts the theory into canonical
form. There are the usual Hamiltonian and momentum constraints associated 
with diffeomorphism invariance, and a new Gauss' law constraint associated
with gauge transformations. Following a standard procedure for
quantizing a system with constraints, one attempts to quantize the theory by
requiring that the constraint operators annihilate the physical states.
The theory is analogous to an SU(2) Yang-Mills theory with an unusual
Hamiltonian. However, an important difference with ordinary Yang-Mills theory
is that there is no background spacetime metric. One must quantize the
theory in a diffeomorphism invariant way. It turns out that there is a
natural diffeomorphism invariant measure on the space of 
connections\footnote{More precisely, there is a diffeomorphism invariant measure
on the space of generalized connections mod gauge transformations. A
generalized connection is a map from line segments (edges) to the group.}
which
can be used to define an inner product. 
It is sometimes convenient to work in a loop representation
in which states  are functionals of loops. The relation to the connection
representation is roughly $\tilde \psi[\gamma] = \int DA \ W_\gamma[A]\ \psi[A]$
where $\gamma$ is a loop and $W_\gamma[A] =Tr P \exp{\oint_\gamma A}$ is
the Wilson loop.

The current status of the quantum geometry approach is roughly the following:

\begin{enumerate}

\item There is a detailed and well defined theory of quantum geometry with
a Hilbert space of states, and operators describing geometrical quantities
such as areas, volumes, etc. This can be viewed as the kinematical Hilbert
space for the gravitational field. (The Hamiltonian constraint has not yet
been imposed.)  It turns out that the fundamental excitations of the
geometry are one dimensional, created by gravitational analogues of Wilson
loops. An orthonormal basis for this space is
given by spin networks, which are, roughly, graphs whose edges are labeled with 
representations of $SU(2)$.

\item Progress has been made in understanding the dynamics, i.e., the
Hamiltonian constraint. (This is much more difficult than the other constraints
since it is quadratic in momenta and needs to be regulated.)
For example, a rigorously defined and finite
Hamiltonian operator has been constructed and states have been found
which are annihilated by this operator \cite{THI}. This defines a
consistent generally covariant four dimensional quantum field theory,
but it may not
reduce to general relativity in the classical limit \cite{MALE} . 
Also there has been
progress in a path integral approach, in which the spin networks are
generalized to ``spin foams" which are two dimensional surfaces glued together 
along edges \cite{RERO}. 

\item There has been recent progress in constructing kinematical
semi-classical states \cite{THM}.
These are quantum states which approximate classical
spacetimes with minimum uncertainty. They are important for gaining a physical interpretation
of quantum states, and are needed as a first step to studying scattering
in this approach.

\item Although supersymmetry does not seem to be necessary in this approach, 
there has been progress in extending this approach to supergravity
\cite{SMO}.

\end{enumerate}

Let me now describe some of the results in the quantum geometry approach.

A) {\it Discreteness:} The geometric operators describing areas and volumes
have a purely discrete spectrum. For example, the area of a small surface
crossing
an edge of a spin network is directly related to the $SU(2)$ representation
on the edge. The area of a large surface is just the sum of the contributions
from each edge of the spin network it crosses.
Since all geometric operators have a discrete spectrum, geometry is really
quantized. The usual continuum picture is only a coarse grained approximation.
This is another answer to the question of what replaces our usual notions
of space and time, if they are not fundamental.

B) {\it Black hole entropy:} One can reproduce the entropy of all nonrotating
black holes (including Schwarzschild) by counting quantum states \cite{ABCK}.
The answer
agrees with the Bekenstein-Hawking prediction up to a single undetermined
dimensionless constant, called the Immirzi parameter. This parameter
arises since there are inequivalent ways of quantizing the classical
phase space \cite{ROTH}.
If you fix this parameter to give the
right answer for Schwarzschild, one automatically gets the right answer
for all charged black holes. This is different from the situation 
in string theory, in which  current calculations
only gives the entropy of a general black hole
up to an undetermined
factor of order one which is not simply related to a parameter in the theory.

One might wonder how one can count physical 
states of the black hole if one does not yet have complete
control over the Hamiltonian constraint. The answer is the following.
If one starts with
the Einstein action in appropriate variables defined outside a stationary
black hole, one can show that one needs to add a surface term at the horizon
which is essentially a $U(1)$ Chern-Simons action (recall that we are in
$3+1$ dimensions). Physically, this Chern-Simons action describes fluctuations
of the horizon geometry. The Hilbert space consists of  states in the
bulk  and states on the horizon, coupled in a well defined way. What one
actually counts are states of the boundary Chern-Simons theory, on a sphere
with certain punctures. The horizon area is kept fixed and corresponds to 
a sum of contributions associated with each puncture.
One then  assumes that each of these boundary 
states can be connected to a bulk state satisfying the Hamiltonian constraint.

C) {\it Area eigenvalues are consistent with Hawking radiation:}
In this framework, Hawking radiation can be thought of as analogous to atomic
transitions: the area drops to a lower eigenstate, and one emits a quanta of
energy. But to reproduce an approximately thermal spectrum at low frequency,
it is important
that the area eigenvalues are not evenly spaced \cite{BEMU}. This can easily
be seen as follows. Suppose the area eigenvalues were $A_n \sim n$ in
Planck units. Even though the Planck length is so small, this would lead
to observable effects. Since $A \sim M^2$, $\delta A \sim M \delta M$. So
$\delta M \sim 1/M = \omega_0$. Thus black hole radiation could only consist
of particles with energy $\omega_0, \ 2\omega_0, \ 3\omega_0$, etc. But
the Hawking temperature of the black hole is of order $\omega_0$, so this
should be the peak of the thermal spectrum. Fortunately, one finds that
the level spacing between area eigenvalues goes to zero very rapidly,
$A_n - A_{n - 1} \sim e^{-\sqrt A_n}$, which is consistent with a thermal
spectrum even at low frequency.

The quantum geometry
approach has recently made contact with another approach to quantizing
general relativity, based on the close analogy with topological field theory.
The basic idea in the following.
Consider a theory in $D$ dimensions
of a gauge field $A_\mu^a$ with gauge group $G$
and $D-2$ form taking values in the (dual of the)
adjoint representation of $G$, $B_{\mu\cdots \nu a}$. The action is simply
$S_{BF} = \int F \wedge B$ where $F$ is the usual field strength of $A$.
This is a topological field theory, which is independent of a spacetime
metric. It is known in the literature as BF theory \cite{BF}.
This theory can be 
quantized by path integral or canonical methods. For $D=3$ and
$G= SO(2,1)$, this action is precisely $2+1$ general relativity, where $B$
is interpreted as a triad of orthonormal vectors. Even in
higher dimensions,  general relativity  is equivalent to $S_{BF}$
plus a constraint.
This can be seen by writing the Einstein action in first order form
\be
S = \int R^{ab} \wedge e^c\cdots \wedge e^d \epsilon_{abc\cdots d}
\ee
where $R^{ab}$ is the curvature two form of an $SO(D-1,1)$ connection 
and $e^a$ is a 
set of $D$ orthonormal vectors.
This is clearly equivalent to $S_{BF}$
with gauge group $G = SO(D-1,1)$ and $B$ 
constrained to take the form of the wedge product of vectors. 
In any dimension, this constraint can be written as a
quadratic condition on $B$ \cite{FRKR}. So general relativity is equivalent to
a topological field with a simple quadratic constraint!


Returning to four dimensions, a functional integral approach
to quantization has
been developed starting with a simplicial decomposition of the four manifold
$M$. The two form $B$ is defined on the faces, and the connection is defined
on the edges of a dual triangulation, in which each i-simplex is replaced
by a (4-i)-simplex. Evaluating the functional integral involves summing over
group representations, and the constraint on $B$ can be simply implemented
by restricting the group representations one must sum over \cite{BACR}. The dual
triangulation involves two dimensional faces joined at edges, and one
can show that this leads to the same description 
as the ``spin foam" model mentioned earlier  \cite{BAEZ}. 
The spin foam model was originally
obtained in a completely different way,
by extending the spin networks  of the canonical theory
to a four dimensional framework. A topological theory can be completely
described by a single triangulation, but general relativity requires a sum
over triangulations. A field theory formulation has been found in which 
the Feynman diagrams are in one-to-one correspondence with the spin foams
\cite{DFKR}.
So summing over Feynman diagrams naturally sums over spin foams, which is like
summing over four geometries. This is analogous to the matrix theory description
of two dimensional gravity. Although much of the work in this direction has
been restricted to Euclidean general relativity (where the group is compact),
there has been recent work in extending this to Lorentzian general relativity.

\section{Comparisons}

 From the mid 1980's to the mid 1990's, a key difference between string 
theory and quantum geometry was that string theory was inherently perturbative
and quantum geometry was not. (Indeed, the latter
was often called ``nonperturbative
quantum gravity" to distinguish it from string theory.) As I have tried
to emphasize, this is no longer the case. There is now a complete 
nonperturbative formulation of string theory (at least for certain boundary
conditions).

Both string theory and quantum geometry have given strong evidence that they
satisfy the black hole constraint: They can reproduce the entropy of black holes
by counting quantum states. But they do so in very different ways.
Quantum geometry is directly counting fluctuations of the event horizon,
while string theory extrapolates the black hole to weak coupling and
counts states of strings (and branes) in flat spacetime. At the moment,
the string calculations give exact results (including the factor of $1/4$)
only for extreme and near extreme charged black holes. Rotation can be
included. In contrast, the quantum geometry calculations apply to 
all nonrotating black holes, even those which are far from extremality. 
They do not recover the factor $1/4$ uniquely, but it can be reproduced by fixing the
one free
parameter in the theory. 
In this approach, one counts horizon states and
assumes that they can be extended to  states satisfying the Hamiltonian 
constraint.

The main advantage of the quantum geometry approach is that it is directly
facing the challenge of quantizing space and time. String theory has not
yet done this. Initially it avoided the issue by focusing on perturbation 
theory. More recently, it has side-stepped the issue by using duality
to relate strong coupling to weak coupling, or using holography to 
remove spacetime altogether. We do not yet have a good understanding of
how to recover spacetime from its holographic description.

The main disadvantage of quantum geometry is that it is still trying to 
deal with the Hamiltonian constraint. This has always been the main
difficulty of canonical quantization of general relativity, and
despite enormous progress, it has not yet been resolved.

The main advantage of string theory is that it is much more ambitious. It
attempts to provide not just a quantum theory of gravity but also a unified
theory of all particles and forces.  String theory has also
achieved more, including the results on singularities and topology change
mentioned earlier, and  more detailed calculations of Hawking radiation.

The main disadvantage of string theory is that one has to accept a lot 
of extra structure: extra dimensions, supersymmetry, extra particles.
At first sight it does not seem very economical. But nature may in fact
include all of these features, and it is only experimental limitations which
have kept them hidden. Furthermore, we have seen evidence that string theory
has fewer degrees of freedom than it appears, so it may be more economical
than it looks.

\section{Looking ahead}

In the next few years, I expect to see progress in both approaches. But in
the long run, things depend on the following key question:

\vskip .5 cm
{\it Does quantum general relativity exist as a consistent theory?}
\vskip .5 cm

If so, the quantum geometry approach will probably succeed and construct it.
If not, it will fail. But even in this case, ideas from quantum geometry
are likely to be useful in string theory e.g. to recover spacetime from
its holographic description.
If quantum general relativity exists and is (in a suitable sense) unique, 
then quantum geometry
must be included in string theory since string theory includes general
relativity.

How could quantum geometry be combined with string theory?
There are already several similarities between the two approaches.
 For example, one dimensional objects play a key role in both.
In string theory, this is the starting point: all elementary 
particles are excitations of a one dimensional string. In quantum geometry,
one finds (at the end of a lengthy analysis) that fundamental excitations
of the geometry are one dimensional.
Given the AdS/CFT conjecture, there are further similarities arising from
the fact that gauge theories play an important role in both approaches.
String theory
with $AdS_5\times S^5$ boundary conditions is completely described by a 
four dimensional $SU(N)$ gauge theory. In the quantum geometry approach,
four dimensional general relativity is described by something resembling an
$SU(2)$ gauge theory. Could the fact that $SU(2)$ is contained in $SU(N)$
be related to the fact that string theory describes many more fields than
just four dimensional general relativity? Furthermore, in the quantum 
geometry approach, Wilson loops are the basic operators creating fundamental
one dimensional excitations of the geometry. In string theory, Wilson loops in
the boundary gauge theory describe strings in the bulk spacetime.

Of course, at the moment there are also crucial differences between these
approaches. The gauge theory in string theory is a standard Yang-Mills theory
with its usual Hamiltonian and a fixed spacetime metric. In quantum geometry,
there is a Hamiltonian constraint, spatial diffeomorphism constraints
and no background metric. Another key difference is that
one of the main predictions of the quantum
geometry approach is that all areas are discrete. But in string
theory we have seen that one can wrap extended objects around compact 
surfaces in the extra
dimensions to produce new states. The mass of these states is directly 
proportional to the area of the surface. Supersymmetry requires that one
can change this area continuously.  So it appears that area is not
quantized in this case. 
The future will tell whether, at a deeper level, these differences
are superficial or fundamental.

Of course the goal of both approaches is to answer basic questions such
as  what was
physics like at the big bang. Given the recent progress, one may be
hopeful that answers will be available soon into the next millennium.

\vskip 1cm
\centerline{\bf Acknowledgements}
\vskip .2cm
It is a pleasure to thank A. Ashtekar, T. Banks, K. Krasnov, D. Marolf,
J. Maldacena, J. Polchinski, C. Rovelli, and A. Strominger
for improving my understanding of the
topics discussed here. 
This work was supported in part by NSF Grant 
PHY-0070895.
\vfill\eject


\begin{thebibliography}{99}

\bibitem{BEK} J. Bekenstein, ``Black Holes and Entropy",
Phys. Rev. D7 (1973) 2333.

\bibitem{HAW} S. Hawking, ``Particle Creation by Black Holes",
Commun. Math. Phys. 43 (1975) 199.

\bibitem{HOMY} G. Horowitz and R. Myers, ``The Value of Singularities",
Gen. Rel. Grav. 27 (1995) 915,
gr-qc/9503062.

\bibitem{GAST} D. Garfinkle and A. Strominger, `` Semiclassical Wheeler
Wormhole Production",
Phys. Lett. B256 (1991) 146.

\bibitem{GGS}  D. Garfinkle, S. Giddings, and A. Strominger, 
``Entropy in Black Hole Pair Production", Phys. Rev. D49
(1994) 958, gr-qc/9306023.

\bibitem{POL} J. Polchinski, {\it String Theory}, in 2 vols., Cambridge Univ.
Press, 1998.

\bibitem{ADD} N.~Arkani-Hamed, S.~Dimopoulos, and G.~Dvali, 
``The Hierarchy Problem and New Dimensions at a Millimeter", Phys. Lett.
 B429 (1998) 263, hep-ph/9803315.

\bibitem{RS} L.~Randall and R.~Sundrum, ``An Alternative to Compactification",
Phys. Rev. Lett.   83 (1999) 4690, hep-th/9906064.

\bibitem{AGM} P. Aspinwall, B. Greene, and D. Morrison,
``Calabi-Yau Moduli Space, Mirror Manifolds and Spacetime Topology Change
in String Theory",
 Nucl. Phys. B416 (1994) 414,
hep-th/9309097.

\bibitem{STR} A. Strominger, ``Massless Black Holes and Conifolds in 
String Theory", Nucl. Phys. B451 (1995) 96, hep-th/9504090.

\bibitem{AMS} B. Greene, D. Morrison, and A.
Strominger, ``Black Hole Condensation and the Unification of String Vacua",
Nucl. Phys. B451 (1995) 109.

\bibitem{STVA} A. Strominger and C. Vafa, ``Microscopic Origin of the
Bekenstein-Hawking Entropy", Phys. Lett. B379 (1996) 99,
hep-th/9601029.

\bibitem{REVS} A. Peet, ``TASI Lectures on Black Holes in String Theory",
hep-th/0008241.

\bibitem{HOPO} L. Susskind, ``Some Speculations about Black Hole
Entropy in String Theory", hep-th/9309145;
G. Horowitz and J. Polchinski, 
``A Correspondence Principle for Black Holes and Strings", 
Phys. Rev. D55 (1997) 6189, hep-th/9612146;
T. Damour and G. Veneziano, 
``Self-gravitating Fundamental Strings and Black Holes", Nucl. Phys. B568 (2000) 93, hep-th/9907030.


\bibitem{MAL} J. Maldacena, ``The Large N Limit of Superconformal Field 
Theories and Supergravity",  Adv. Theor. Phys.  2 (1998) 231,
hep-th/9711200.

\bibitem{MAGOO} O. Aharony, S. Gubser, J. Maldacena, H. Ooguri, and Y. Oz,
``Large N Field Theories, String Theory and Gravity", 
 Phys. Rept.  323 (2000) 183, hep-th/9905111.

\bibitem{HOO}  G. 't Hooft, ``Dimensional Reduction in Quantum Gravity",
gr-qc/9310026.

\bibitem{SUS}  L. Susskind, ``The World as a Hologram", J. Math.
Phys. 36 (1995) 6377, hep-th/9409089.

\bibitem{ROV} C. Rovelli, ``Loop Quantum Gravity", Living Reviews 1 (1998),
gr-qc/9710008.

\bibitem{THI} T. Thiemann, 
``Anomaly-free Formulation of Non-perturbative, 
Four-dimensional Lorentzian Quantum Gravity",
Phys. Lett. B380 (1996) 257, gr-qc/9606088.

\bibitem{MALE} J. Lewandowski and D. Marolf, ``Loop Constraints: A Habitat 
and their Algebra",  Int. J. Mod. Phys. D7 (1998) 299,
gr-qc/9710016.

\bibitem{RERO} M. Reisenberger and C. Rovelli, ``Sum over Surfaces form of
Loop Quantum Gravity", Phys. Rev. D56 (1997) 3490,
gr-qc/9612035.

\bibitem{THM} T. Thiemann, ``Gauge Field Theory Coherent States",
hep-th/0005233.

\bibitem{SMO}  Y. Ling and L. Smolin, 
``Supersymmetric Spin Networks and Quantum Supergravity", Phys.
Rev. D61 (2000) 044008, hep-th/9904016; ``Holographic Formulation of 
Quantum Supergravity", hep-th/0009018.

\bibitem{ABCK} A. Ashtekar, J. Baez, A.
Corichi  and K. Krasnov, ``Quantum Geometry and Black Hole Entropy",
Phys. Rev. Lett. 80 (1998) 904, gr-qc/9710007;
A. Ashtekar, J. Baez, and K. Krasnov, ``Quantum Geometry of Isolated 
Horizons and Black Hole Entropy", gr-qc/0005126.


\bibitem{ROTH} C. Rovelli and T. Thiemann, 
``The Immirzi Parameter in Quantum General Relativity", 
Phys. Rev. D57 (1998) 1009.

\bibitem{BEMU} J. Bekenstein and V. Mukhanov, ``Spectroscopy of the 
Quantum Black Hole", Phys. Lett. B360 (1995) 7,
gr-qc/9505012.

\bibitem{BF} G. Horowitz, ``Exactly Soluble Diffeomorphism Invariant
Theories", Commun. Math. Phys. 125 (1989) 417; M. Blau and G. Thompson,
``Topological Gauge Theories of Antisymmetric Tensor Fields", Ann. Phys.
205 (1991) 130.

\bibitem{FRKR} L. Freidel, K. Krasnov, R. Puzio, ``BF Description of
Higher-Dimensional Gravity Theories", hep-th/9901069.

\bibitem{BACR} J. Barrett and L. Crane, 
``Relativistic Spin Networks and Quantum Gravity",
J. Math. Phys. 39 (1998) 3296, gr-qc/9709028.

\bibitem{BAEZ} J. Baez, ``An Introduction to Spin Foam Models of 
Quantum Gravity and
BF Theory", gr-qc/9905087; 
J. Barrett, ``State Sum Models and Quantum Gravity",
gr-qc/0010050.

\bibitem{DFKR} R. De Pietri, L. Freidel,
K. Krasnov, and C. Rovelli, ``Barrett-Crane Model from a Boulatov-Ooguri
Field Theory over a
Homogeneous Space", Nucl. Phys. B574 (2000) 785, hep-th/9907154;  
A. Perez, ``Finiteness of a Spinfoam Model for Euclidean Quantum General
Relativity", gr-qc/0011058. 

\end{thebibliography}
\end{document}